\pdfoutput=1

\documentclass[sigconf]{acmart}

\AtBeginDocument{%
  \providecommand\BibTeX{{%
    \normalfont B\kern-0.5em{\scshape i\kern-0.25em b}\kern-0.8em\TeX}}}

\copyrightyear{2023}
\acmYear{2023}
\setcopyright{rightsretained}
\acmConference[SC '23]{The International Conference for High Performance Computing, Networking, Storage and Analysis}{November 12--17, 2023}{Denver, CO, USA}
\acmBooktitle{The International Conference for High Performance Computing, Networking, Storage and Analysis (SC '23), November 12--17, 2023, Denver, CO, USA}
\acmDOI{10.1145/3581784.3607073}
\acmISBN{979-8-4007-0109-2/23/11}

\usepackage{subfigure}
\usepackage{subcaption}
\usepackage{pdfpages}
\usepackage{import}
\usepackage{graphicx}

\begin{document}

\title{Hanayo: Harnessing Wave-like Pipeline Parallelism for Enhanced Large Model Training Efficiency}

\author{Ziming Liu}
\authornote{Both authors contributed equally to this research.}
\email{liuziming@comp.nus.edu.sg}
\affiliation{%
\country{}
  \institution{National University of Singapore}
}
\author{Shenggan Cheng}
\authornotemark[1]
\email{shenggan@comp.nus.edu.sg}
\affiliation{%
\country{}
  \institution{National University of Singapore}
}

\author{Haotian Zhou}
\email{zhou0000@comp.nus.edu.sg}
\affiliation{%
\country{}
  \institution{National University of Singapore}
}

\author{Yang You}
\email{youy@comp.nus.edu.sg}
\affiliation{%
\country{}
  \institution{National University of Singapore}
}

\begin{abstract}
   Large-scale language models have become increasingly challenging and expensive to train. Among various methods addressing this issue, Pipeline Parallelism has been widely employed to accommodate massive model weights within limited GPU memory. This paper introduces Hanayo, a wave-like pipeline parallelism strategy that boasts a concise structure and practical applicability, alongside a high-performance pipeline execution runtime to tackle the challenges of pipeline strategy implementation. Hanayo mitigates the issues of pipeline bubbles and excessive memory consumption prevalent in existing schemes, without resorting to model duplicates as in Chimera. Our evaluation, conducted on four distinct computing clusters and involving both GPT-like and BERT-like architectures with up to 32 GPUs, demonstrates up to a 30.4 \% increase in throughput compared to the state-of-the-art approach.
\end{abstract}





\maketitle

\section{Introduction}

Over the past decade, deep learning has made significant strides in numerous fields, including Computer Vision (CV) and Natural Language Processing (NLP). Among various architectures, the transformer \cite{vaswani2017attention} has emerged as a prominent model due to its exceptional sequence modeling capabilities. Recent studies have demonstrated that transformers not only surpass recurrent neural networks in NLP \cite{devlin2018bert, raffel2019exploring} but also outperform many convolutional neural networks in CV \cite{dosovitskiy2020image, liu2021swin}. It has been shown that substantial performance gains can be attained by utilizing large-scale datasets in conjunction with expansive transformer-based models.

In the past six years, the number of model parameters has increased 40-fold every 18 months, while in the last three years, it has grown 340-fold within the same period. Currently, models can contain hundreds of billions of parameters \cite{brown2020language}. This rapid increase in model parameters outpaces the memory expansion of accelerators, necessitating the use of large-scale GPU supercomputing clusters for training. As a result, the time and financial costs of training large models have escalated, becoming almost prohibitive, as exemplified by the Megatron-Turing NLG 530B model \cite{smith2022using} developed by Microsoft and NVIDIA, which required approximately three months to train on over 2,000 A100 GPUs.

Several challenges arise from the continuous growth in model sizes, including: 1) \textit{Memory Wall}, where the model's parameter size significantly exceeds the storage capacity of a single accelerator, often by several orders of magnitude; 2) \textit{Scaling Wall}, which arises when training large models necessitates the use of thousands of accelerators, resulting in complex parallel patterns and extensive communication that can lead to bottlenecks in scaling; 3) \textit{Computational Wall}, referring to the immense computational power demanded by large models and massive datasets; and 4) \textit{Development Wall}, where the intricate parallel strategies and manual control of communication processes render the development of large model training exceedingly difficult.

Facing the above challenges, the mainstream approach for training large models involves employing model parallelism techniques. In contrast to data parallelism, where each device contains a full set of model parameters, model parallelism distributes the parameters across different devices. There are two primary model parallelism methods: tensor parallelism and pipeline parallelism.

Pipeline parallelism focuses on parallelization at the layer level, with layers assigned to different devices. While tensor parallelism is associated with significant communication costs, pipeline parallelism relies on peer-to-peer communication for transferring intermediate activations, resulting in considerably lower communication overhead. This makes pipeline parallelism an indispensable strategy for large model training. For instance, in the parallel strategy employed by Megatron-LM \cite{narayanan2021efficient}, tensor parallelism is utilized within nodes (intra-node), while pipeline parallelism is applied across nodes (inter-node).

Nonetheless, the current pipeline parallelism approach has several limitations that diminish the overall efficiency of large model training: 1) \textit{Bubble}, which refers to the idle time of devices due to computation dependencies in the pipeline across different devices; 2) \textit{Communication Overhead}, even though pipeline parallelism employs point-to-point (P2P) communication, its overhead remains considerable; 3) \textit{Memory Consumption}, some pipeline schemes attempt to mitigate the bubble issue by relying on model replication\cite{li2021chimera}. However, this approach exacerbates the already stringent GPU memory constraints, which might further complicate the training of large models; 4) \textit{Hybrid Programming Paradigm}, since pipeline parallelism is incompatible with the prevalent SPMD (single program, multiple data) programming paradigm for distributed training, it poses challenges for flexible pipeline process scheduling within modern deep learning frameworks.

To address the aforementioned challenges, we introduce Hanayo, a unified pipeline parallelism framework featuring a wave-like pipeline scheme. 1)To break the memory wall, Hanayo decouples the reduction of bubble ratio with model replication, requiring the same level or lower memory compared with mainstream methods; 2)For scaling, the Hanayo unified framework enables the expression of mainstream pipeline parallel algorithms in a universal manner while facilitating further generalization that enables us to automatically scale pipelines to more devices; 3)For computation, we lower the overall bubble ratio with our unique pipeline scheme, wave pipeline, reducing the idles and achieving high throughput with the same computing power; 4)For easier development, we orchestrate the training process using an \textit{action list}, allowing for support of virtually all pipeline parallel algorithms within the runtime system. Besides, we have also implemented efficient communication schemes with techniques such as pre-fetching.

In summary, our paper presents the following contributions:

\begin{itemize}
\item We introduce a wave-like pipeline scheme that achieves a low bubble ratio and high performance in large model training. It can achieve increasingly higher throughput as the number of waves increases.
\item Hanayo proposes a unified framework for pipeline parallelism. Through theoretical analysis, we obtain a unified performance model for pipeline parallelism. 
\item In the design and implementation of the runtime system, we aim to decouple the runtime system from specific pipeline parallel algorithms. Utilizing the action list, Hanayo's runtime system can support nearly all pipeline parallel algorithms while optimizing performance through features such as asynchronous communication.
\item We conduct experiments with mainstream GPT-style and BERT-style models, performing performance tests for various model sizes on four different computing clusters. Experimental results demonstrate that Hanayo achieves up to a 30.4\% performance improvement over the current state-of-the-art pipeline parallelism implementation, Chimera.
\end{itemize}


\section{Background}

During deep neural networks (DNN) training, the main computations are divided into two processes:  \textit{forward propagation} (FP) and  \textit{backward propagation} (BP) . In forward propagation, we calculate the loss of different samples operated by the model using the objective function. In backward propagation, we backpropagate the loss using the chain rule of differentiation to calculate gradients, which are used to update the parameters\cite{lecun2015deep, rumelhart1985learning}.

\begin{figure}[hbt] 
\centering 
\includegraphics[width=0.44\textwidth]{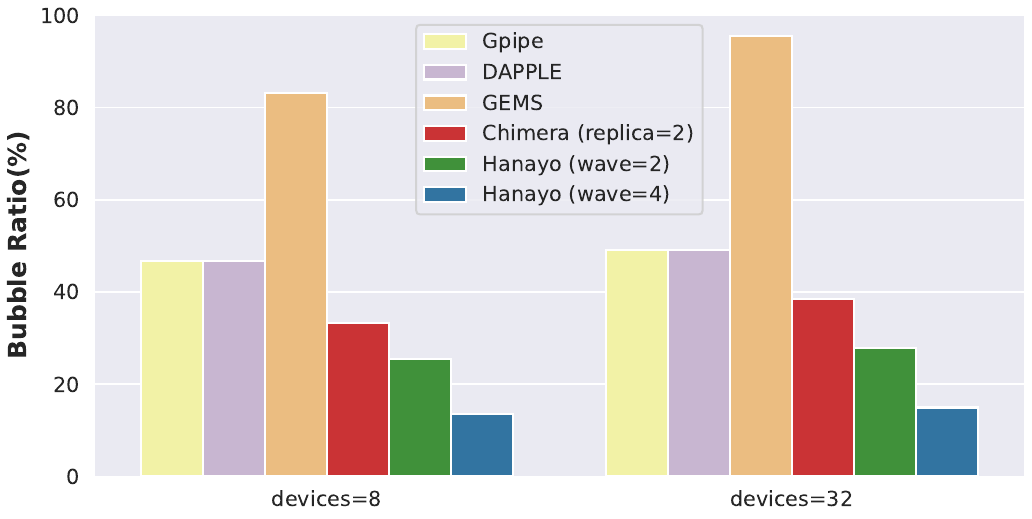} 
\caption{The theoretical bubble ratio of synchronous pipeline schemes} 
\label{Fig.bubble_rate}
\end{figure}

\subsection{Parallelism Technique for Training}

As DNN training data is often massive, for example, GPT-3 was trained with 45 TB of text data\cite{wei2021finetuned}. Therefore, we need to use \textit{Data Parallelism} (DP) \cite{hillis1986data, shoeybi2019megatron, li2020taming, you2018imagenet} to accelerate DNN training. In forward propagation, each device has a complete and identical model for computing. In backward propagation, each device computes gradients and synchronizes them through all-reduce to update the parameters.

However, as the size of DNN models grows, it becomes impractical to train them using data parallelism alone due to the limited memory on each device. Model parallelism has been introduced to address this issue. Model parallelism distributes the model parameters to different devices to achieve lower memory consumption. Based on the different ways of dividing the model parameters, we can generally categorize Model Parallelism into \textit{Tensor Parallelism} (TP)\cite{shoeybi2019megatron, wang2022tesseract, xu2021efficient} and \textit{Pipeline Parallelism} (PP)\cite{gpipe, dapple, li2021chimera, yang2021pipemare, pipedream, yang2022groupbased}. 


Tensor parallelism means splitting a tensor into chunks along a specific dimension and each device only holds a part of the whole tensor while not affecting the correctness of the computation graph. It involves distributing an operator's parameters to different devices. Each device then computes a local result based on its assigned data slices. Finally, at the end of the operator computation, collective communication (such as all-gather or all-reduce) is inserted to obtain the final result. Despite its advantages, tensor parallelism comes with a higher communication overhead because of the need for synchronizing communication after each split tensor operation. This effect is particularly pronounced during cross-node training, where the low communication bandwidth can significantly reduce the training speed.

To reduce the communication volume between nodes, pipeline parallelism has been proposed. Pipeline parallelism partitions the model at the layer level, while also partitioning the mini-batch into micro-batches. Each worker is treated as a pipeline stage for the forward propagation or backward propagation computation of a micro-batch.

\subsection{Synchronous Pipeline Parallelism}
To enable a more comprehensive analysis of the pros and cons of state-of-the-art pipeline methods, we have standardized the symbolic representation, which is presented in Table \ref{tab.symbo}. In this section, we will focus on the major synchronous pipeline parallelism algorithms, such as GPipe \cite{gpipe}, DAPPLE \cite{dapple}, and Chimera \cite{li2021chimera}, and we will also introduce some asynchronous approaches.

\begin{table}[hbt]
\caption{Meanings of the symbols that are used in this paper}
\centering
\renewcommand\arraystretch{1.2}
\label{tab.symbo}
\begin{tabular}{ll}
\noalign{\hrule height 1pt}
S    & The number of pipeline stages                                            \\
B    & The number of micro-batches in a single iteration               \\
D    & The number of replicated pipelines                                       \\
P    & The number of workers used in the pipeline                               \\
W    & The number of waves in a single forward/backward\\
 & iteration(=S / (2 * P)) \\
$M_w$ & Memory consumption for the weights of one stage                          \\
$M_a$ & Memory consumption for the activation of one stage                      \\
$T_F$ & Time cost for a complete forward pass\\ &(all forward stages added together) divided by P                      \\
$T_B$ & Time cost for a complete backward pass\\ &(all backward stages added together) divided by P                      \\
$T_C$ & Time cost for a single P2P communication\\ \noalign{\hrule height 1pt}
\end{tabular}
\end{table}

\begin{figure*}[hbt]
  \vspace{-3mm}
  \subfigure[GPipe] {
    \label{fig:GPipe}
    \begin{minipage}[b]{0.85\textwidth}
    \includegraphics[width=1\columnwidth]{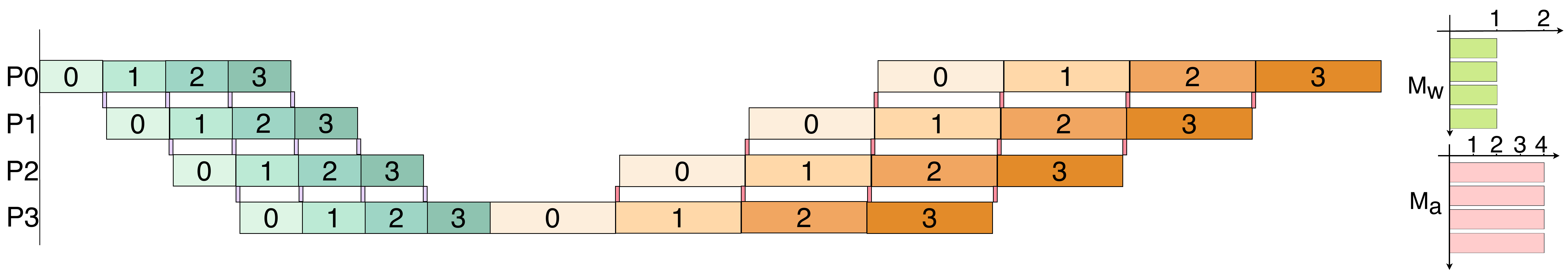}
    \end{minipage}
  }    
  \vspace{-3mm}
  \subfigure[DAPPLE] { 
    \label{fig:DAPPLE}
    \vspace{-10mm}
    \begin{minipage}[b]{0.85\textwidth}
    \includegraphics[width=1\columnwidth]{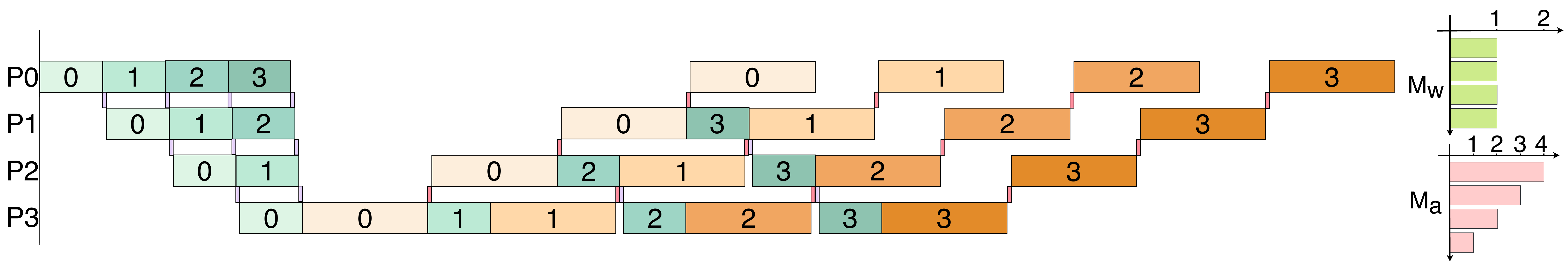}
    \end{minipage}
  }
  \vspace{-3mm}
  \subfigure[Chimera] {
    \label{fig:chimera}  
    \begin{minipage}[b]{0.85\textwidth}
    \includegraphics[width=1\columnwidth]{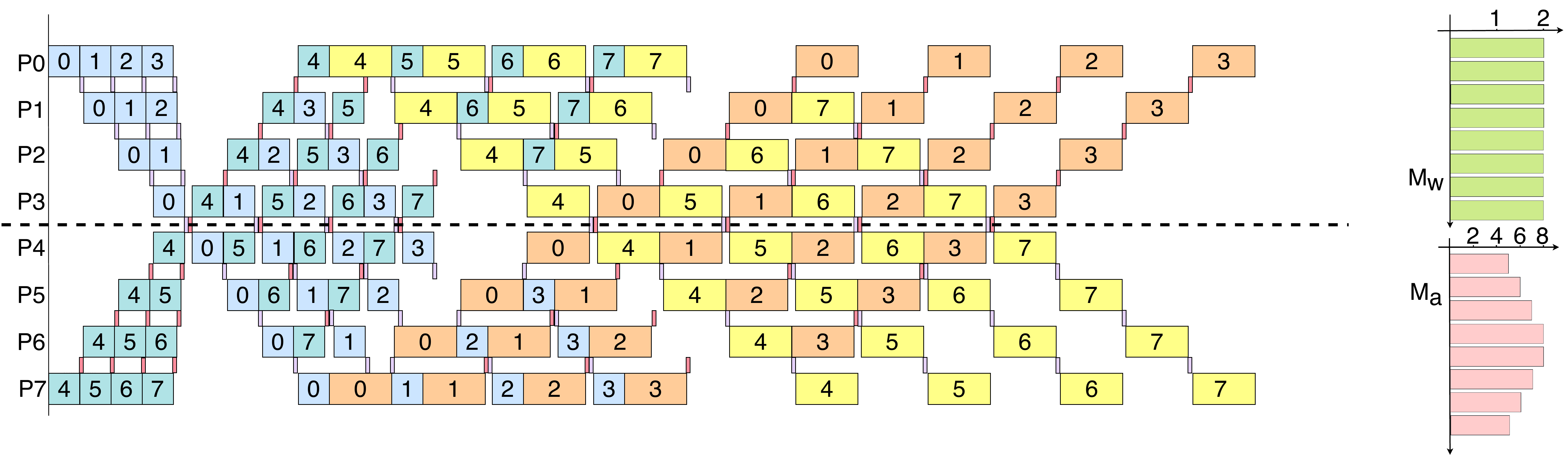}
    \end{minipage}
  }
  \vspace{-3mm}
  \subfigure[Hanayo with one wave] {
    \label{fig:wpipe_1}    
    \begin{minipage}[b]{0.85\textwidth}
    \includegraphics[width=1\columnwidth]{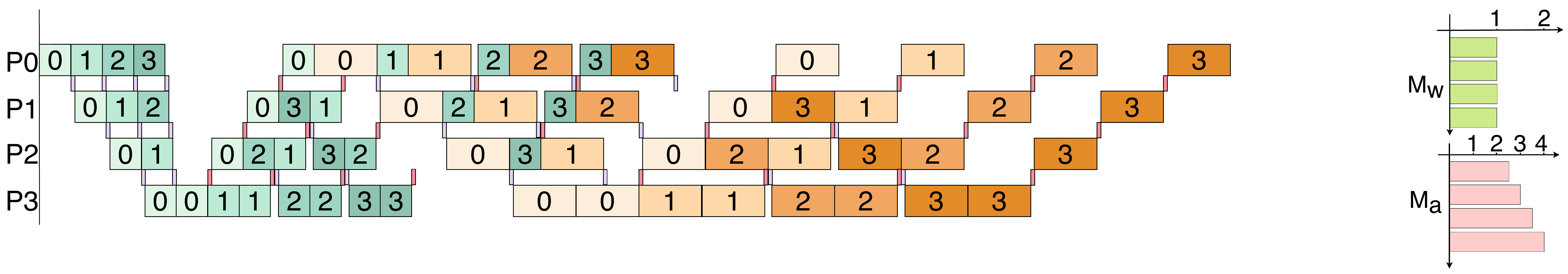}
    \end{minipage}
  }
  \vspace{-3mm}
  \subfigure[Hanayo with two waves] {
    \label{fig:wpipe_2}     
    \begin{minipage}[b]{0.85\textwidth}
    \includegraphics[width=1\columnwidth]{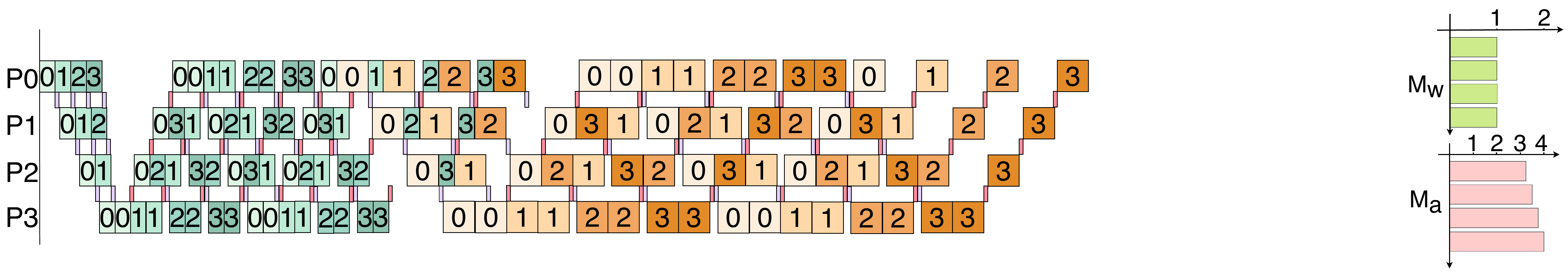}
    \end{minipage}
  }
  \caption{Synchronous Algorithms of Pipeline Parallelism and their peak memory consumption. The forward propagation process is marked green and the backward propagation process is marked orange. Blue and yellow are also used in Chimera to mark the two directions in it. Back propagation is illustrated twice as long as forward propagation according to the training experience. And we use purple and pink blocks to represent the P2P communication that goes downward and upward. The numbers on the blocks show which of the micro-batches is being processed. Each unit block in $M_w$ represents a whole model weight divided by the number of devices, and one unit block in $M_a$ represents one intermediate activation.}     
  \label{fig:sync_pipeline}
\end{figure*}

The most classic pipeline parallel algorithm is GPipe, as illustrated in Figure \ref{fig:GPipe}. In this example, the minibatch is divided into four micro-batches for the four devices shown in the figure. The four devices pipeline the forward computation of the four micro-batches and then pipeline the backward computation of the four micro-batches after the completion of all forward computations. The pipeline is divided into three main parts. The first part is at the beginning of the forward computation, where a device must wait for the previous device to complete the calculation of its corresponding microbatch. The second part is between the forward and backward computations, where a device is waiting for the corresponding microbatch to be calculated after completing the forward computation. The third part is after the backward computation is finished, where the device that finished the backward computation must wait for the flush. GPipe is a relatively simple and efficient pipeline parallel algorithm. However, it requires all intermediate activations of the microbatch to be saved during training, which results in relatively high memory consumption.

DAPPLE utilizes the One-Forward-One-Backward (1F1B) schedule to enhance the GPipe, making it one of the most widely used pipeline parallelism methods in practical applications. As shown in Figure \ref{fig:DAPPLE}, 1F1B adjusts the forward and backward order of different microbatches on different devices. By calculating the backward earlier, 1F1B releases the intermediate activation stored on the device as early as possible, providing it with a certain memory advantage over GPipe. However, we have observed that the memory consumption of 1F1B is uneven across devices: the peak memory usage does not drop on the first device, while it can significantly decrease on the last device. And Megatron-LM has proposed some improvements to the 1F1B algorithm by introducing interleaving, which divides the model and data more finely, creating more opportunities for overlapping and thus reducing the bubble ratio.

Chimera introduces a novel bidirectional pipeline parallelism technique that achieves a smaller bubble ratio and is currently one of the most effective methods for pipeline parallelism, with more balanced memory consumption. In Figure \ref{fig:chimera}, we show a typical Chimera pipeline with 2 model replicas on 8 devices, using blue and yellow blocks to mark the forward and backward propagation going downward, and green and orange for those going upward. By storing another slice of the model on each device, a worker can do computation going upward while waiting for the activation from the pipeline going downward. The two pipelines can thus fill in each other's bubbles. In this way, Chimera exhibits a lower bubble ratio compared to prior pipeline methods, indicating higher theoretical training efficiency. Furthermore, the use of bidirectional pipeline reduces memory consumption imbalances across different cards. However, due to its support for bidirectional pipeline, Chimera needs to store a duplicate copy of model parameters in both directions, resulting in twice the memory overhead of other methods. Further analysis and discussion of Chimera can be found in the next section.

\begin{figure*}[hbt] 
\centering 
\includegraphics[width=0.9\textwidth]{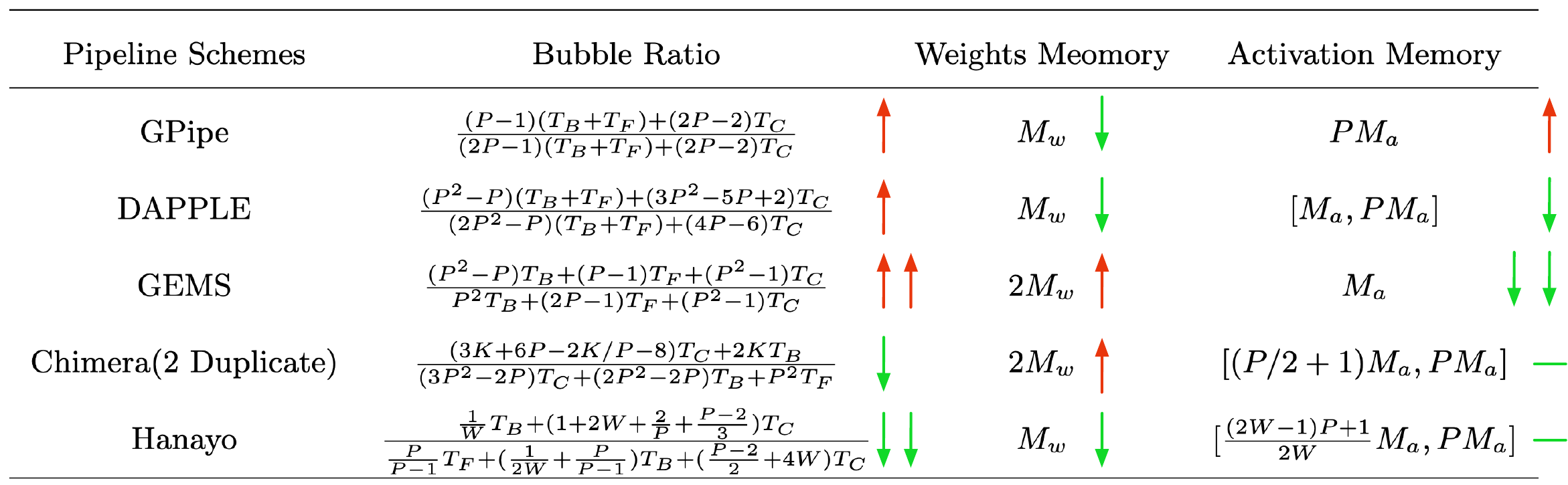} 
\caption{Comparison of different SOTA approaches \\ Green arrows indicate better performance} 
\label{Fig.formula}
\end{figure*}

In Figure \ref{Fig.formula}, we present a comparative analysis of different approaches with respect to GPU memory consumption and bubble ratio, highlighting their respective advantages and disadvantages. Notably, we factor in the impact of remaining communication time after overlap when calculating the bubble ratio, resulting in a more accurate assessment. The up red arrow in Figure \ref{Fig.formula} denotes a higher ratio or consumption, which implies inferior performance. Conversely, the down green arrow signifies a lower ratio or consumption, indicating superior performance. K in bubble ratio of Chimera is defined as $K=\frac{P^2}{2} - P$.

\subsection{Asynchronous Pipeline Parallelism}

The synchronous pipeline parallelism algorithm performs a flush at each computation step, allowing the training process to maintain consistency by using the same version of model parameters for each batch. To achieve a lower bubble ratio, recent works have proposed asynchronous pipeline parallelism algorithms, such as PipeDream \cite{pipedream}, WPipe \cite{yang2022groupbased}, and PipeMare \cite{yang2021pipemare}.

\begin{figure}[hbt]
  \centering
  \subfigure[Synchronous Pipeline Parallelism] {
    \label{fig:sync}     
    \includegraphics[width=0.80\columnwidth]{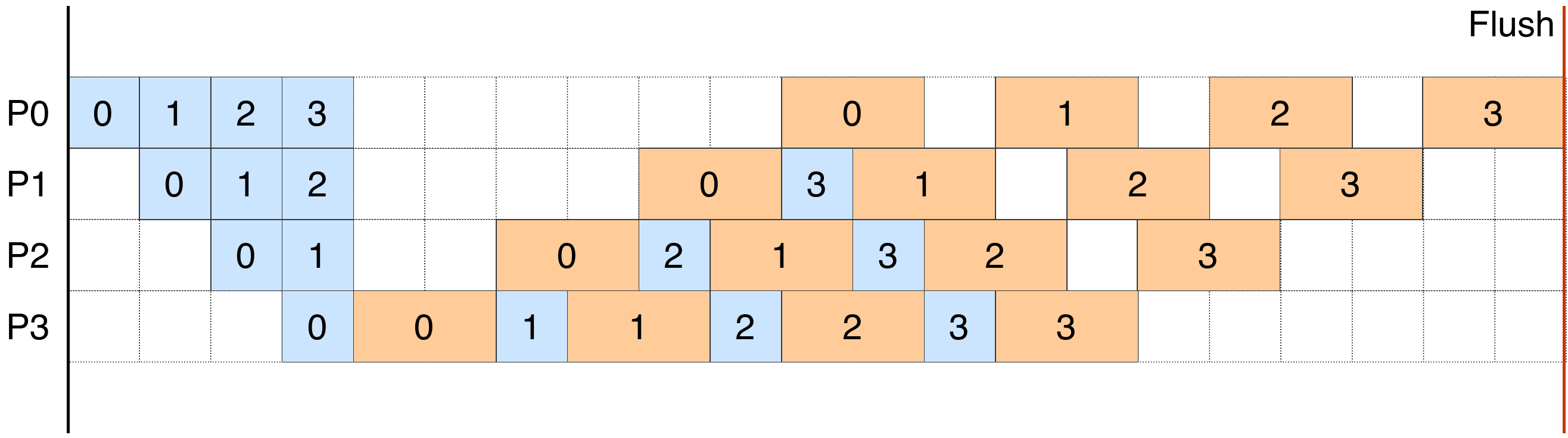}  
  }    
    
  \subfigure[Asynchronous Pipeline Parallelism] { 
    \label{fig:async}     
    \includegraphics[width=0.80\columnwidth]{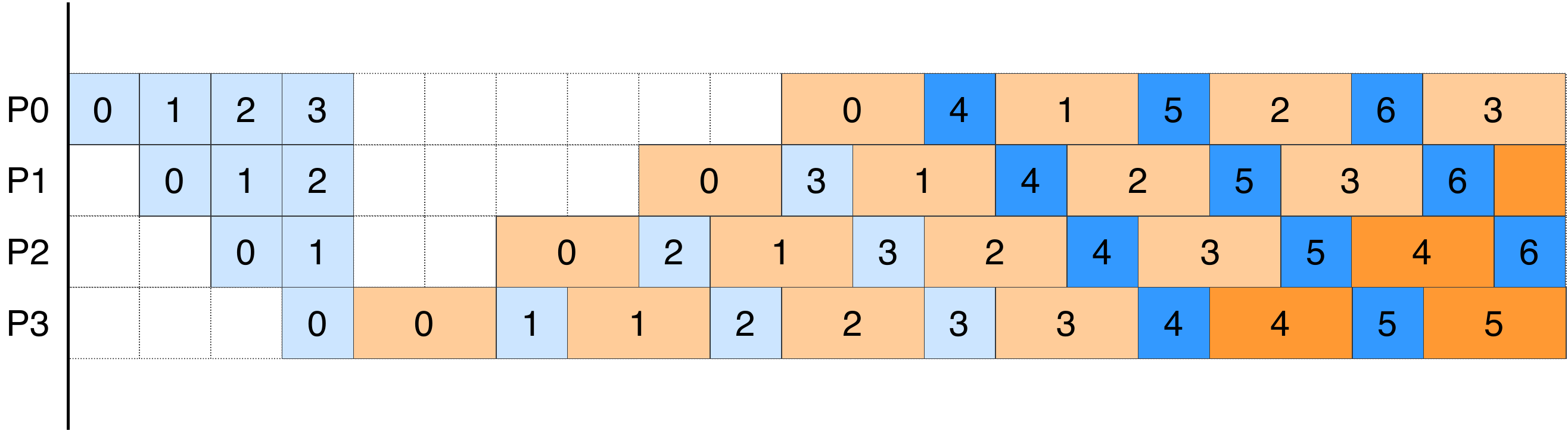}
  }    
  \vspace{-10pt}
  \caption{Comparison of synchronous and asynchronous Pipeline Parallelism}     
  \label{fig:sync_async}
\end{figure}

Unlike synchronous pipeline parallelism, asynchronous approaches remove the flush and allow for more relaxed dependency constraints. As a result, they tend to have a lower bubble ratio and higher performance, as illustrated in Figure \ref{fig:sync_async}. However, asynchronous optimization can impact the convergence process of training. Although there have been works to improve asynchronous convergence\cite{Zhang2016ASGD}, they often come with extra computation or memory overhead. So we focus only on synchronous pipeline parallelism in this work. Nevertheless, the strategies and optimizations we propose can also be applied to asynchronous pipeline parallelism implementation.

\section{HANAYO UNIFIED FRAMEWORK}

\subsection{Motivation}
As previously discussed, pipeline parallelism faces two significant challenges, namely memory consumption and computation efficiency. While GPipe provides a simple and easy-to-implement solution, it suffers from high activation consumption and a high bubble ratio. DAPPLE addresses the memory consumption challenge by modifying the computation schedule, resulting in lower memory consumption. However, it still struggles with a high bubble ratio. Chimera proposes a promising solution with its bi-directional pipeline approach that orchestrates multiple pipelines and achieves an impressive low bubble ratio. But Chimera requires model replicas, which may not be feasible for limited GPU memory or large models. This leads us to consider an alternative approach: can we rearrange the computation scheme so that we can benefit from the low bubble ratio without extra memory consumption?

\subsection{Transforming into Wave-like Pipelines} \label{transforming}
We have found a simple way that leads us out of this dilemma. We know that the high efficiency of Chimera can be primarily attributed to its bidirectional pipeline structure, allowing pipelines in different directions to compensate for bubbles. The reason for employing model replication is that, in the current pipeline scheduling, the same micro-batch must continuously perform calculations and communication in the same direction. Therefore, when introducing calculations in another direction, another set of models must be stored on the GPU. To address this issue, we only need to enable a single pipeline to change direction during the computation process, transforming it into a wavy-shaped pipeline.

\begin{figure}[hbt]
\centering 
\includegraphics[width=0.44\textwidth]{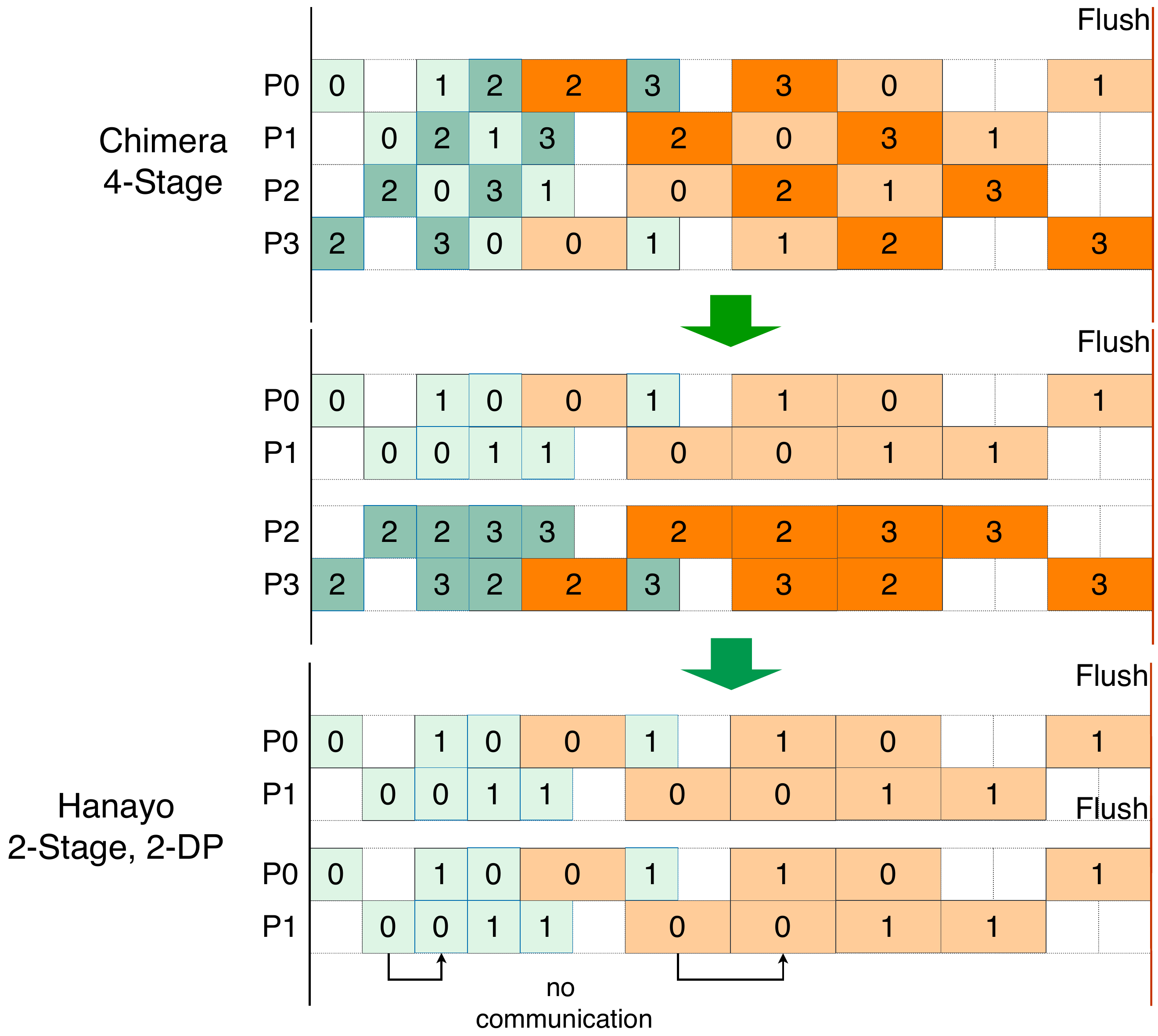} 
\caption{A 4-stage Chimera pipeline can be transformed into two one-wave pipelines with a 2-stage Data Parallel without extra overhead.} 
\label{Fig.swap}
\end{figure}

Here we have a typical bi-directional Chimera pipeline with a depth of 4, as illustrated in Figure \ref{Fig.swap}. The darker and brighter blocks represent computations performed by the two pipelines with different directions, which we will refer to as $Pipe_{bright}$ and $Pipe_{dark}$, respectively. In this specific case, micro-batch 0 and 1 are $Pipe_{bright}$ as their forward propagation goes downward, while micro-batch 2 and 3 are $Pipe_{dark}$ as their forward propagation goes upward. As depicted in the figure, if we swap all the $Pipe_{bright}$ blocks on devices P2 and P3 with the corresponding $Pipe_{dark}$ blocks located at symmetrical positions on devices P0 and P1, we obtain two identical wave-like pipeline structures. Since the order of computation remains unchanged and the communication overhead is reduced due to the local communication introduced by this swap operation, we can infer that the efficiency of these two wave-like pipelines is at least as good as, if not better than, the original 4-stage Chimera configuration. By adopting this approach, we eliminate the need for model replication in the pipeline implementation, and the replicas employed by Chimera can now be considered as standard data parallelism. There is also an example in figure\ref{fig:chimera} and figure\ref{fig:wpipe_1}. 

Why we call such pipeline scheme wavy may not be quite obvious in this case, so we have another example in figure \ref{fig:w2d8}. We mark the whole training process of micro-batch 1 with the color blue and red. There are four "V"s in the whole training process for each micro-batch, which look exactly like waves. In this paper, we define the number of "V"s in a forward or backward propagation process the number of waves. So we would say that there are two waves on 8 devices in figure \ref{fig:w2d8}.

To ensure a fair and convenient comparison, we will measure Chimera after transforming it into its corresponding wave-like form, which we will call Chimera-wave in this paper. As Chimera-wave can optimize Chimera by reducing cross-communication, it enables us to view model duplication as an expansion of the data parallelism scale, allowing us to measure Chimera while consuming the same amount of memory for model weights as other methods.

\subsection{More waves For Lower Bubble Ratio}
Having successfully eliminated model duplication, we can now focus on reducing the bubble ratio. During the calculation of the bubble ratio for current pipeline schemes, we observe that with a fixed number of devices, the factors that influence the bubble ratio (or the total idle time in the pipeline) are $T_F$, $T_B$, and $T_C$. As $T_C$ is fixed in a given cluster environment, we can investigate the remaining factors, which represent the time cost of a forward stage and a backward stage. By partitioning the model into smaller stages, we can achieve improved performance.

In the wavy pipeline scheme derived from Chimera, the number of stages is already twice as much as that of regular pipelines. What if we continue doubling it? This can be easily achieved by incorporating more waves into the pipeline. In Figure \ref{fig:wpipe_2}, the number of stages increases from eight to sixteen as the number of waves rises from one to two. It is evident that the sizes of all the bubbles resulting from waiting for peer devices' computation are halved, while the total computation remains unchanged. Although the number of times that we do communication is now doubled, most of them can be overlapped by the computation, and the additional bubbles caused by cross-communication are more than compensated for by the overhead saved by smaller bubble size, yielding a significantly lower bubble ratio.

\begin{figure*}[hbt]
    \subfigure[wave=2, devices=8] {
        \label{fig:w2d8}
        \begin{minipage}[t]{1\textwidth}
            \includegraphics[width=1\columnwidth]{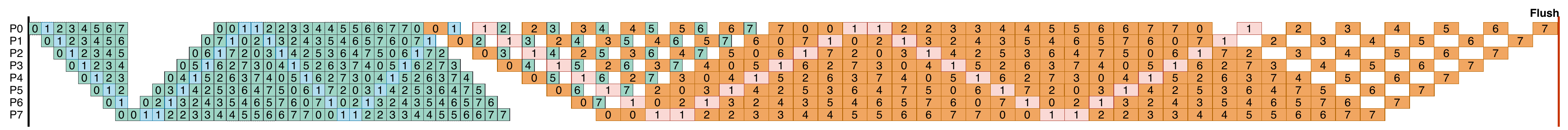}
        \end{minipage}
    }    

    \subfigure[wave=2 and wave=4, devices=4] { 
        \label{fig:d4}     
        \begin{minipage}[t]{0.5\textwidth}
            \includegraphics[width=\textwidth]{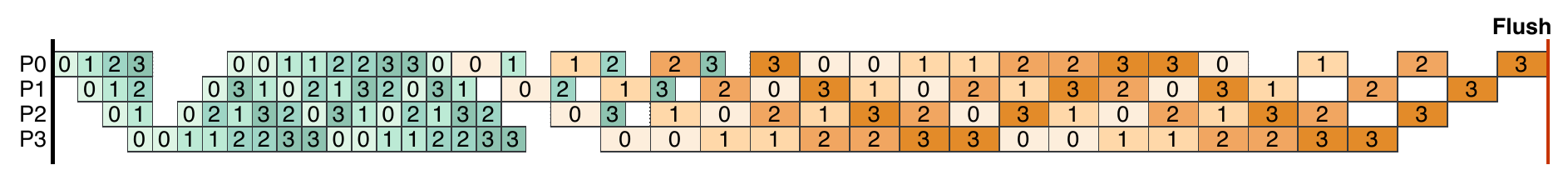}
        \end{minipage}\hfill
        \begin{minipage}[t]{0.5\textwidth}
            \includegraphics[width=\textwidth]{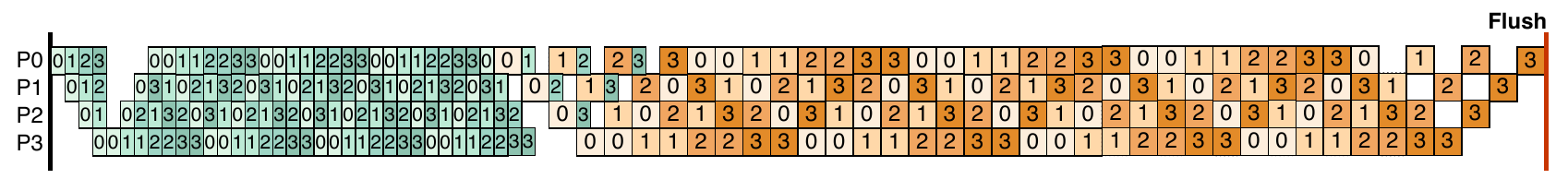}
        \end{minipage}
    }
  \caption{Scaling Hanayo to more devices and waves}     
  \label{fig:scale_wpipe}
\end{figure*}

Furthermore, we can continue to increase the number of waves as long as there are sufficient layers within a single stage to divide. In Figure \ref{fig:d4}, we present an example where we expand the number of waves to four, resulting in the bubble sizes being halved once more. The implementation is quite straightforward, as the structure of the intermediate waves remains identical. Similarly, Hanayo can be scaled to accommodate more devices by employing the corresponding number of mini-batches during the training process. In Figure \ref{fig:w2d8}, we implement Hanayo on 8 devices with the number of waves set to two. We have highlighted micro-batch 1 to illustrate the direction of a single computation process. The large number of stages may appear confusing; however, there is no need to grapple with the structure, as we have developed an ingenious pipeline framework that can automatically deploy the structure with any desired number of waves or devices. We will discuss this framework in greater detail later.

\begin{figure}[hbt] 
\centering 
\includegraphics[width=0.44\textwidth]{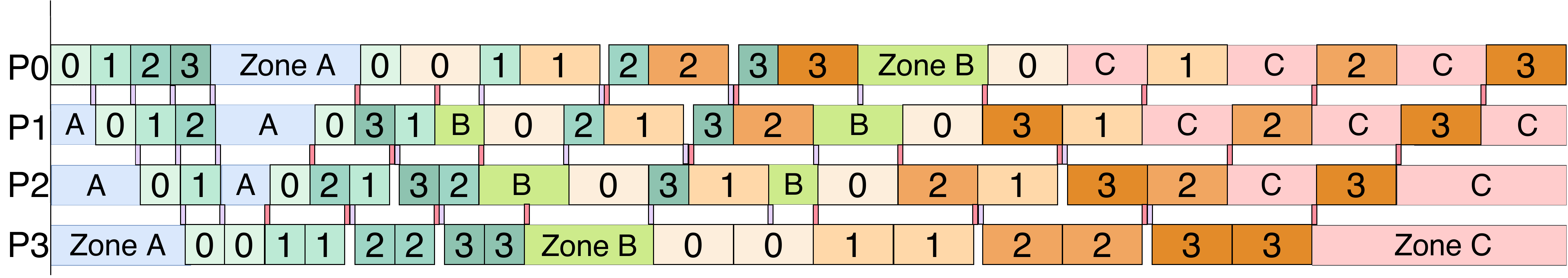} 
\caption{The type of bubbles that exist in a typical Hanayo wave-like pipeline} 
\label{Fig.wpipe_theory}
\end{figure}

\subsection{Theoretical Analysis}
In this section, we delve into the bubbles that impact the performance of Hanayo. There are four distinct types of bubbles in a training iteration, as illustrated in Figure \ref{Fig.wpipe_theory}. The bubbles in Zone A primarily result from idle waiting for forward activation from peer workers and the overhead of forward activation transmission. Thus, the size of a single bubble in this zone is given by $T_F / 2W + T_C$, as the forward time cost is reduced by a factor of $2W$. The bubbles in Zone B mainly arise from the discrepancy between the overheads of forward and backward propagations. Consequently, the bubble size in B is calculated as $\frac{P-LR}{2W}(T_B-T_F) + 2T_C$, where LR represents the local rank. As for bubble type C, it is caused by backward propagations and the corresponding communication, resulting in bubble sizes of $T_B + 2T_C$ and $T_B + T_C$. Lastly, bubbles due to cross-communication occur when using the NCCL\cite{nccl} backend, which requires batching cross-communication together before initiation to prevent deadlock. By summing all four types of bubbles, we can compute the final bubble ratio:
\begin{equation}
    \frac{\frac{1}{W}T_B + (1+2W+\frac{2}{P}+\frac{P-2}{3})T_C}{\frac{P}{P-1}T_F + (\frac{1}{2W} + \frac{P}{P-1})T_B + (\frac{P-2}{2} + 4W)T_C}
\end{equation}
Assuming $T_B=2T_F$ and disregarding $T_C$, we can rewrite this equation in a more suitable form as $\frac{2P-2}{3PW+P-1}$. This expression decreases with an increasing number of waves, demonstrating the efficiency of such wave-like structures. Moreover, since we do not employ model replicas or alter the computation order, our $M_w$ and $M_a$ remain consistent with those of other mainstream methods. In figure \ref{Fig.bubble_rate}, we show the theoretical bubble ratio for the pipeline schemes. You can see a sharp drop in Hanayo's bubble ratio with an increased number of waves.

\section{Hanayo Runtime}
In this section, we present the design and implementation of our runtime system, which primarily consists of the pipeline execution engine decoupled from the pipeline parallelism scheduling algorithm and communication optimization, including prefetching and asynchronous communication.

\subsection{Implementation Scheme}
Our implementation scheme is inspired by the widely-used distributed deep learning optimization library, DeepSpeed \cite{10.1145/3394486.3406703}. DeepSpeed employs a set of instructions, such as ForwardPass, BackwardPass, and SendActivation, in its schedulers. Workers use an interpreter to read the instructions and execute the corresponding actions. We observe that some of these instructions are not well-suited for more complex structures like Hanayo or Chimera. To address this, we break down the instructions into smaller granularities and augment them with target device rank information and local module rank (indicating which part of the model should be utilized for the given instruction).

Unlike some pipeline frameworks that use hooks to establish the relationship between different stages, our approach employs an \textit{action list} to store the instructions for each worker. The scheduler on the master node is responsible for generating the action list based on a specific pipeline. We have provided scheduling algorithms for existing mainstream pipeline schemes and Hanayo, and we also offer interfaces for users to modify existing schemes or develop their own. 

\subsection{Overlap by Prefetching}
One of the main benefits of using such an action list is that the workers can prefetch the next batch of data from peer devices. Our approach employs the pre-fetching technique with asynchronous communication functions. Before initiating a slice of computation, the processor looks ahead to check the next \texttt{receive} instruction and launches the subsequent \texttt{receive} request before the current forward/backward propagation. This ensures that the activation/gradient for the next round is ready by the end of the current computation round. We know that the transmission of activation and gradients can be costly, especially for large models with high hidden dimensions and computing clusters with poor interconnection. Under such conditions, one key to achieving higher throughput is to maximize the overlap between computation and communication. The prefetching technique employed by Hanayo ensures the least number of bubbles caused by communication and improves overall efficiency. 
In our implementation, we utilized the NCCL\cite{nccl} backend and Pytorch Distributed Library, as the NCCL backend offers better support for GPU operations. Moreover, we use the \texttt{batch\_isend\_irecv} function from the NCCL backend to circumvent deadlock in cross-communication.
\section{Evaluation}
Our experiments were primarily conducted in four environments:
\begin{itemize}
    \item The LONESTAR6 cluster from the Texas Advanced Computing Center (TACC). Lonestar6 comprises 560 compute nodes and 16 GPU nodes. The A100 GPUs have 40 GB of high-bandwidth memory. In each GPU node, there are three GPUs, with GPU 0 on socket 0 and GPU 1 and 2 on socket 1.
    \item A CVM cloud server from Tencent that we rent. The GN10Xp computing node we utilize features an Intel Xeon Cascade Lake 8255C (2.5 GHz) CPU and 8 NVIDIA V100 GPUs with 32GB of memory. The GPUs are connected with NVLink\cite{nvlink}, and the interconnect configuration is described in the V100 architecture whitepaper\cite{nvidiav100}.
    \item A local cluster with 8 NVIDIA A100 GPUs that have 80GB memory. GPU 0 and 1, 2 and 3, 4 and 5, and 6 and 7 are connected with NVLink.
    \item A local cluster with 8 NVIDIA A100 GPUs that have 80GB memory. The GPUs are fully connected with each other with NVLink.
\end{itemize}
The models we use for evaluation include two variants of BERT and GPT-3. The BERT-style model consists of 64 layers, 64 attention heads, and a hidden size of 2560, while the GPT-style model has 128 layers, 16 attention heads, and a hidden size of 1024.

We implemented all mainstream synchronous pipeline schemes, including GPipe, DAPPLE, Chimera, and Hanayo. For each setting, we determined the best parallelism configuration by adjusting the devices used in data parallelism and pipeline parallelism to maximize throughput. As previously mentioned, we evaluate Chimera's performance after transforming it into its wave form to ensure fairness with other methods that only use one set of model weights. The model replicas used by Chimera are considered as additional data parallelism.

In the following section, we focus on addressing the following questions:
\begin{itemize}
\item What is Hanayo's memory consumption like in practice? Is it suitable for GPUs with smaller memory capacities?
\item How adaptable is Hanayo to different computing environments? Can it maintain its superior performance when faced with various computational power and GPU interconnection configurations?
\item How does Hanayo perform in weak scaling settings, where the computing resources and tasks increase simultaneously?
\item How does Hanayo perform in strong scaling settings, where the goal is to use more computing resources to accelerate a fixed?
\end{itemize}

It needs to be mentioned that the Chimera that we compare with in evaluation is the optimized wave version, Chimera-wave, which has better performance than Chimera with 2 model replicas. More details are in section \ref{transforming}.

\begin{figure}[hbt] 
\centering 
\includegraphics[page=1, width=0.4\textwidth]{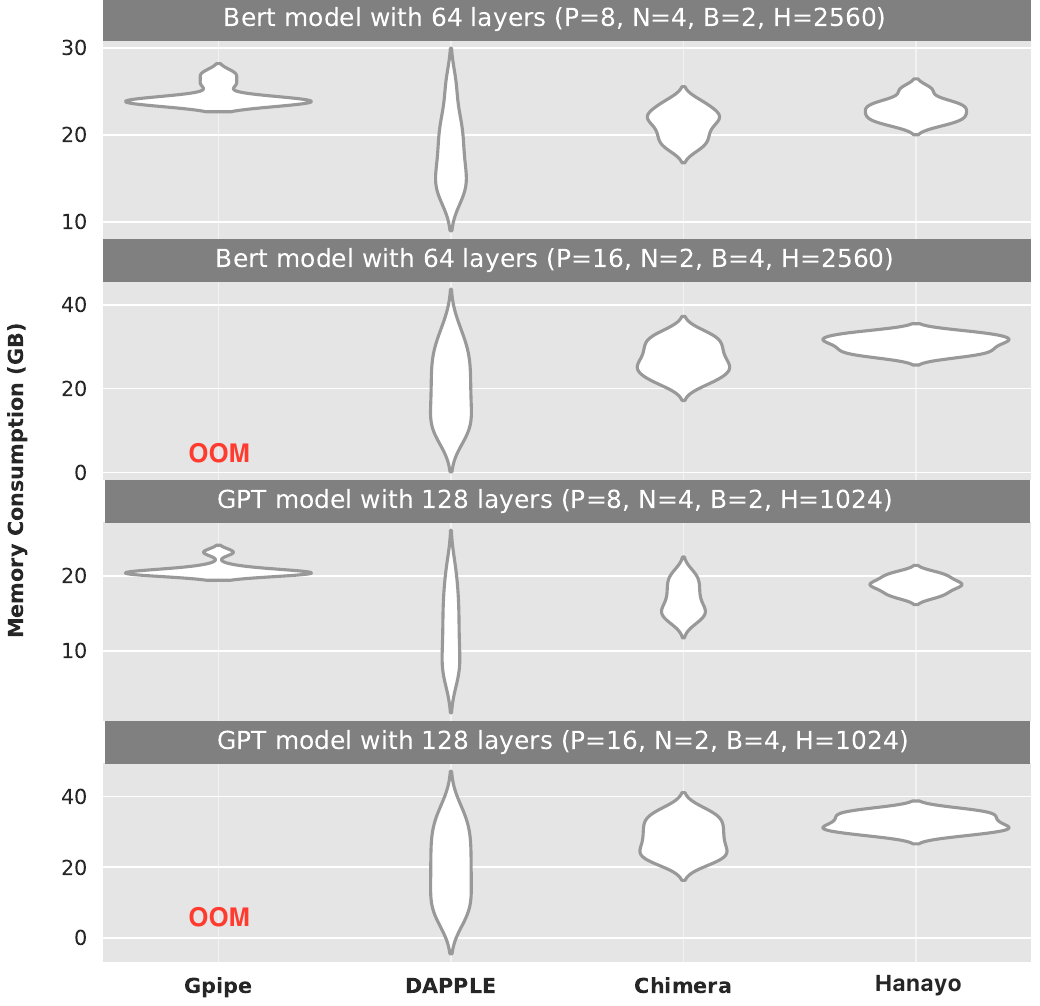} 
\caption{The distribution of peak memory consumption for GPipe, DAPPLE, Chimera, and Hanayo during the training of Bert and GPT model on 32 GPUs of the TACC Lonestar6 cluster} 
\label{Fig.violin}
\end{figure}


\begin{figure*}[hbt] 
\centering 
\includegraphics[width=0.8\textwidth]{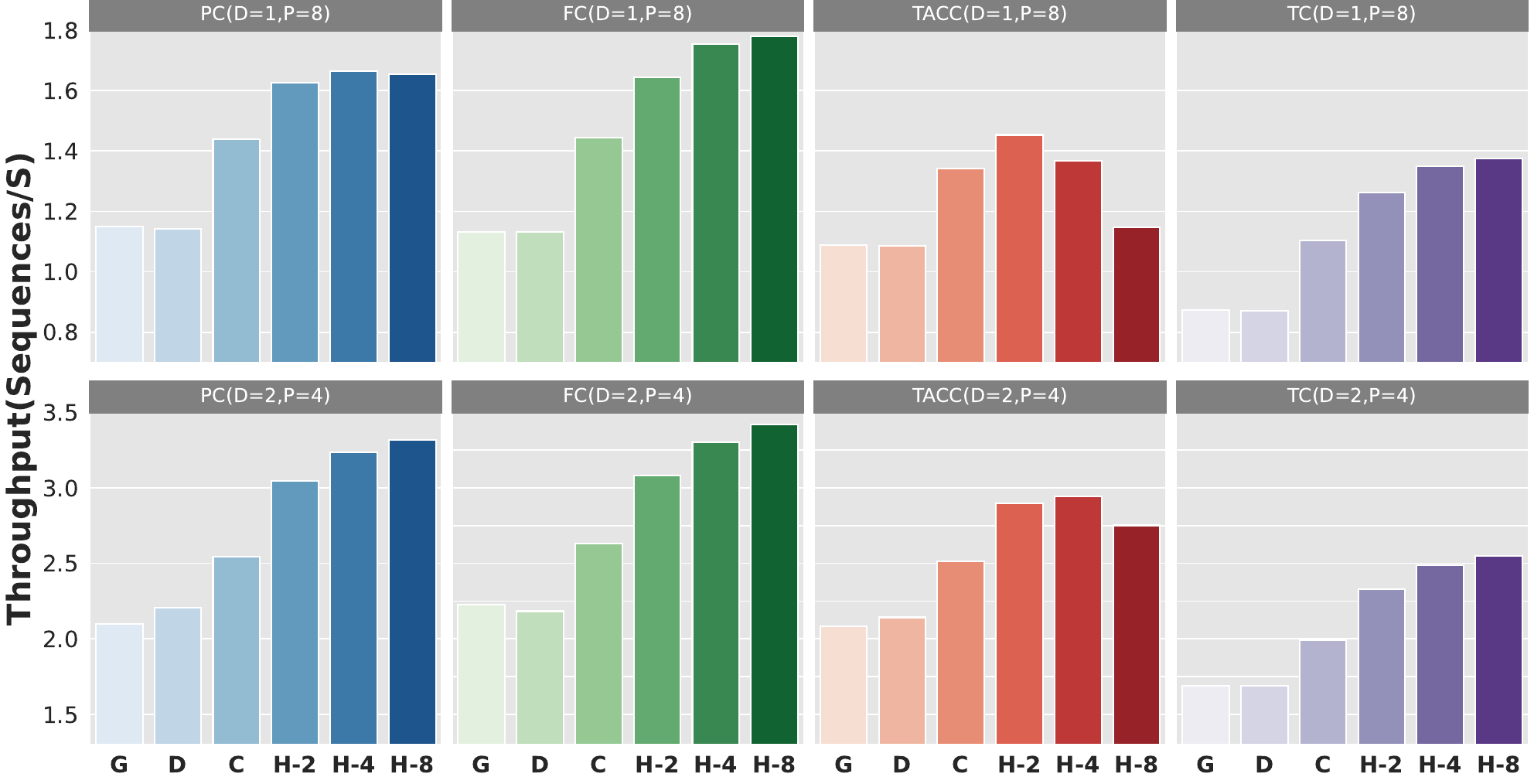} 
\caption{Throughput of training the Bert-style model on totally 32 GPUs from 4 different clusters. PC and FC refer to the two local clusters where the NVIDIA A100 GPUs are partially and fully connected with NVLink. TACC refers to the Lonestar6 cluster from TACC and TC refers to the cloud server of Tencent. As for the methods, G stands for GPipe, D stands for DAPPLE, C stands for Chimera-wave, and H-X stands for Hanayo with X waves.} 
\label{Fig.devices}
\end{figure*}

\subsection{Memory Consumption}

For the two models mentioned earlier, we measured the peak memory distribution across the 32 GPUs utilized during training. The primary factor we need to focus on is the highest peak memory. Although memory consumption varies from device to device, the ability of a scheme to fit within a certain cluster is often determined by the highest peak memory. The computation schemes of GPipe and DAPPLE are relatively similar; however, GPipe requires saving the activation of all micro-batches on every device, resulting in a higher memory consumption for each device. Our findings show that GPipe and DAPPLE have comparable highest peak memory values, but GPipe caused Out of Memory (OOM) errors in two settings. In contrast, Chimera and Hanayo, which benefit from a schedule that consumes the activation as soon as it is generated, achieve lower highest peak memory values.

Another essential aspect to consider is balance. A pipeline scheme with a more balanced memory consumption enables better workload distribution across computational tasks, maximizing the work done on each worker simultaneously and, as a result, achieving higher throughput. GPipe's memory consumption is fairly balanced, with an average variance of $\textbf{1.33}$. However, this low variance comes at the expense of high average consumption. DAPPLE and Chimera do not exceed the memory limit, but their average variances are $\textbf{16.85}$ and $\textbf{2.86}$, respectively, which are higher compared to other methods. Our proposed method, Hanayo, maintains the variance as low as $\textbf{1.44}$ while exhibiting a similar average consumption as Chimera.

In summary, Hanayo exhibits a similar level of peak memory consumption as the state-of-the-art pipeline schemes while maintaining a more balanced memory consumption profile. This balance leads to higher GPU utilization and, ultimately, improved overall performance.

\subsection{Adaptability Across Computing Clusters}

Our first throughput evaluation aims to test the performance of different pipeline schemes in various computing cluster environments. As observed in Figure \ref{fig:sync_pipeline}, the bubble ratio is determined by the communication overhead and the overhead of forward and backward propagation. Despite our efforts to maximize overlap, some communication cannot be covered by computation. In practice, the computation overhead is typically determined by device computing power, while the communication overhead is influenced by various factors such as GPU connections with NVLink\cite{nvlink}, GPU interconnection topological structure, and the bandwidth between computing nodes for cross-node training. In essence, communication overhead is difficult to predict and can vary significantly under different conditions. Therefore, we must evaluate pipeline schemes across diverse environments to assess their adaptability and robustness.

The four selected environments are representative of different use cases. The TACC Lonestar6 computing cluster represents supercomputers or large-scale computing clusters used by companies; the Tencent TVM GPU cloud server represents cloud servers commonly employed for large model training; and the other two clusters are typical local servers used in university labs or small companies. The GPU types include NVIDIA A100 80GB, NVIDIA A100 40GB, and NVIDIA V100 32GB, connected with four distinct topological structures, ensuring diversity in both computation power and communication conditions.

Results are displayed in Figure \ref{Fig.devices}. We limited the number of devices to 8 for this comparison, enabling us to assess performance across clusters. We experimented with two settings: pipeline parallelism only, and a combination of pipeline parallelism size 4 with data parallelism size 2. We did not use a smaller pipeline parallelism size due to the large number of parameters in the BERT model. GPipe and DAPPLE maintain similar throughput across the experiments, as their pipeline stage scheduling primarily differs in activation consumption rather than total time overhead. Chimera outperforms these two methods by approximately 20\% in the eight experiments, owing to its bi-directional design. In contrast, Hanayo offers an additional dimension, the number of waves, which we can adjust in the experiments. We explored all possible wave numbers and selected the best-performing one as our result. In the eight settings, Hanayo consistently outperforms Chimera by $\textbf{15.7\%, 30.4\%, 23.2\%, 29.9\%, 8.2\%, 17.1\%, 24.6\%, and 28.0\%}$, thanks to the wave structures that repeatedly halve the bubble size. This demonstrates Hanayo's advantage over other pipeline parallelism schemes, regardless of the environment.

Another observation from this experiment is that Hanayo's optimal wave configuration can vary with the communication environment. The more waves used, the greater the communication required in one iteration. Although most communication can be covered by computation, there is still more uncovered communication, particularly cross-communication. In servers with full (FC) or partial connections (PC) via NVLink and the Tencent server, which also has NVLink, throughput increases with the number of waves(except for the first setting where the wave number equals 8). This indicates that the improvements brought about by the waves outweigh the additional communication overhead. For clusters with poor interconnection, such as TACC, the optimal wave number will be lower since the extra communication incurs a higher cost.

In conclusion, Hanayo demonstrates superior adaptability across various environments and consistently outperforms state-of-the-art methods. Better topological structures and higher bandwidth further enhance Hanayo's performance.

\subsection{Obtaining The Best Performance For Each Scheme}
\begin{figure}[hbt] 
\centering 
\includegraphics[width=0.4\textwidth]{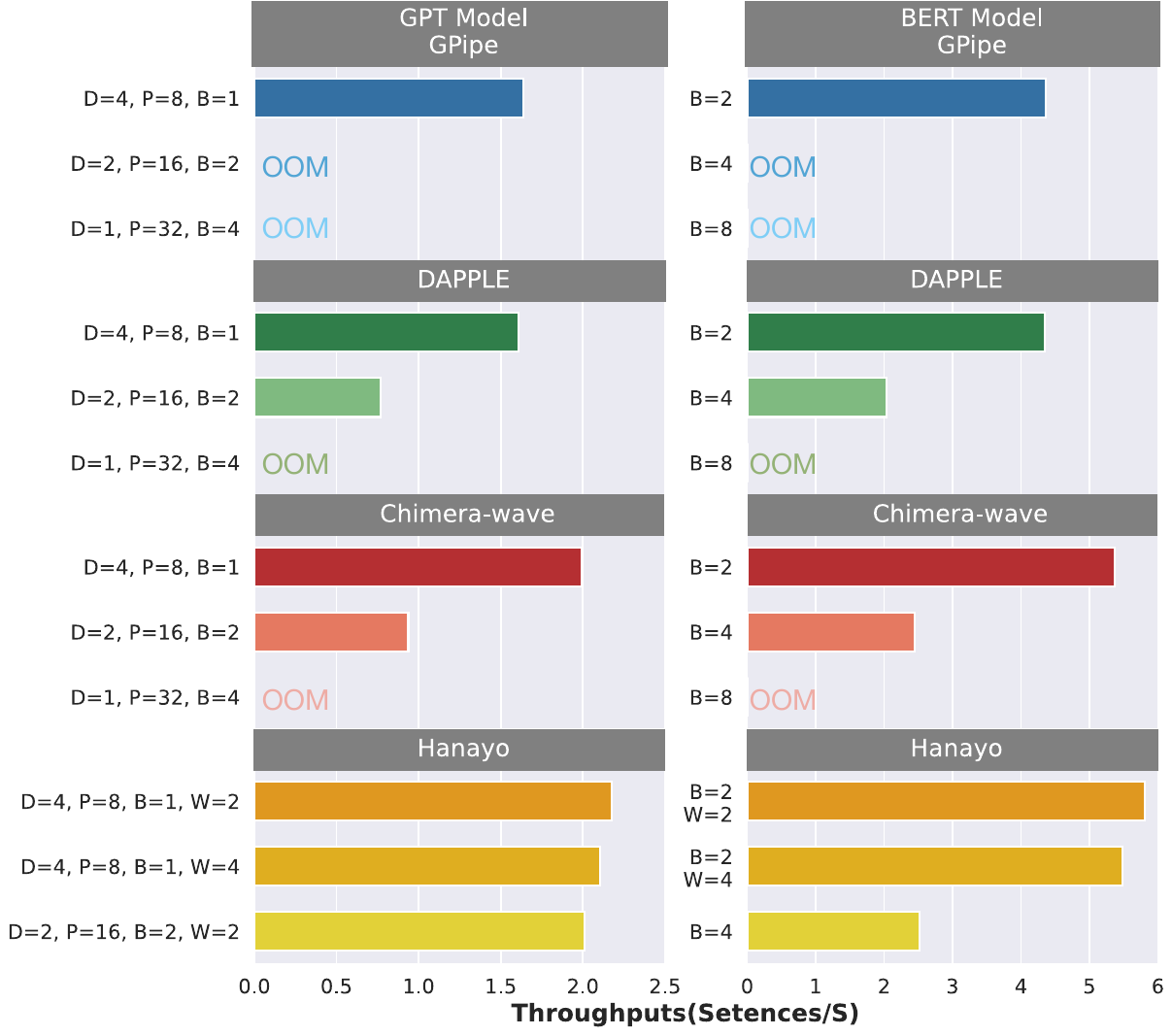} 
\caption{Part of the performance search for the four methods of training the Bert-style model on 32 V100 GPUs from TACC. The configurations with the highest throughput are chosen as targets to be used for further comparison.} 
\label{Fig.best}
\end{figure}
Before delving into scaling, we first describe how we obtain the best throughput data for each pipeline parallelism scheme. In Figure \ref{Fig.best}, we present the search space for each method under the setting where 32 GPUs are used to train the GPT and BERT models mentioned earlier, using the TACC Lonestar6 cluster. The batch size is set to 4 and 8 to maximize GPU memory usage. For each method, we tested the pipeline and data parallelism size combinations of (8, 4), (16, 2), and (32, 1). The absence of data in certain areas indicates that the respective method caused an Out of Memory (OOM) error. It is worth noting that we do not list the wave configuration of Hanayo in this figure. Instead, we searched for the best wave number under each parallelism configuration and displayed them in the corresponding locations. Ultimately, we selected the (D=4, P=8) configuration for all methods, as it yields the highest performance, and chose the number of waves for Hanayo as 2.

\subsection{Weak Scaling}

\begin{figure}[hbt] 
\centering 
\includegraphics[width=0.4\textwidth]{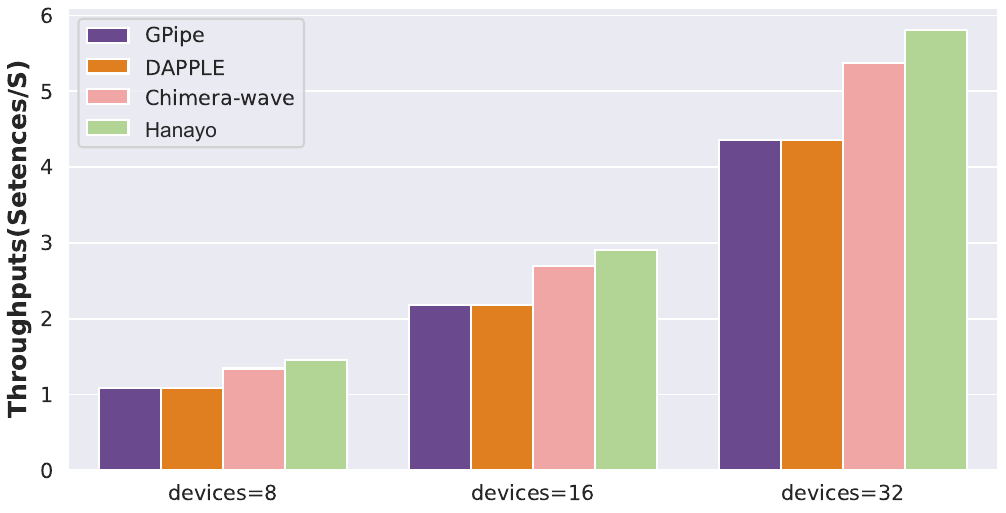} 
\caption{Weak scaling for Bert-style model. The number of devices scales from 8 to 32 while the batch size increases proportionally} 
\label{Fig.weak}
\end{figure}

In this section, we evaluate Hanayo's efficiency under weak scaling settings, where we maintain the same amount of computation per device while incrementally increasing the number of devices used from 8 to 32. The total batch size is increased from 2 to 8. All throughput data were selected using the approach described in the previous section. The results are displayed in Figure \ref{Fig.weak}. In the three sets of experiments, Hanayo outperforms Chimera by $8.19\%, 8.11\%$ and $8.13\%$, DAPPLE by $33.7\%, 33.2\%$ and $33.1\%$, and GPipe by $33.4\%, 33.3\% $ and $33.3\%$. This can be primarily attributed to the reduced bubble size in Hanayo's wave-like structure. Despite the lower bandwidth of GPUs in the TACC Lonestar6 cluster, the advantages of waves still outweigh the extra communication overhead.

Weak scaling is employed to measure a system's ability to maintain computational efficiency as the number of devices and the scale of tasks increase simultaneously. From the results, we can observe that the parallel efficiency is $\textbf{100.1\%}$ and $\textbf{99.8\%}$. The reason why the efficiency exceeds 100\% is that GPUs are better at processing batches of data at the same time thanks to their ability of parallel computing. This demonstrates that Hanayo can be scaled to train larger data batches using larger clusters while maintaining reasonable efficiency.

\subsection{Strong Scaling}

\begin{figure}[hbt] 
\centering 
\includegraphics[width=0.4\textwidth]{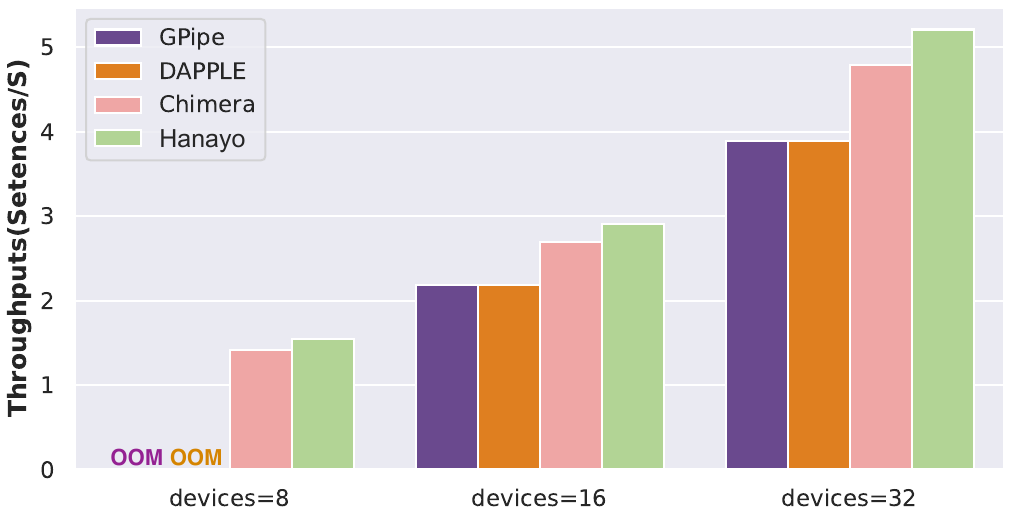} 
\caption{Strong scaling for Bert-style model. We speed up a fixed batch of training with more devices, from 8 to 32.} 
\label{Fig.strong}
\end{figure}

Strong scaling is characterized by maintaining a constant task size while increasing the number of processors. In the context of training a large language model, we attempt to use more GPUs to train the model with a fixed data batch. Here, we use a batch size of 4, which already reaches Lonestar6's 40GB memory limit. We then increase the number of GPUs from 8 to 16 and 32, examining whether the total throughput increases proportionally. The results can be found in Figure \ref{Fig.strong}. GPipe and DAPPLE encounter OOM when using 8 GPUs and achieve similar throughput in the other two cases. Hanayo consistently attains the highest throughput in all three tested cases, outperforming GPipe and DAPPLE by $33.3\%$ and $33.8\%$ and Chimera by $8.8\%,8.1\%$ and $8.7\%$.

Comparing the three sets of data, we can calculate the speedup of Hanayo to be $\textbf{188.4\%}$ and $\textbf{337.5\%}$. This demonstrates its ability to accelerate a specific task with more GPUs. A typical scenario would involve fine-tuning, in which users seek to adjust the released public model weights to achieve better performance on downstream tasks with a small amount of additional training data. As shown in the figure, Hanayo is well-suited to handle this situation.

\section{Related Work}

\textbf{Pipeline Parallelism Scheduling Algorithms.} There has been a surge in recent research focused on large model training, with some of it centering around the pipeline parallelism scheduling algorithms. Among the earliest efforts were GPipe \cite{gpipe} and PipeDream \cite{pipedream}, which remains one of the most widely used Pipeline methods. More recent works such as DAPPLE \cite{dapple} and Chimera \cite{li2021chimera} aim to reduce the bubble ratio in pipeline parallelism by utilizing different scheduling strategies, thereby enhancing the efficiency of large model training. WPipe\cite{yang2022groupbased} proposed a scheme that achieved better
memory efficiency and fresher weight updates in asynchronous pipeline parallelism. The author also posited its potential application in GPipe. Our work builds upon these efforts by presenting a more universal scheduling approach for pipeline parallelism, based on a thorough analysis of state-of-the-art pipeline parallelism methods. In our integrated pipeline framework, we use a performance model with adaptability to choose from various pipeline parallelism strategies to attain optimal performance.

\textbf{Runtime Systems for Pipeline Parallelism.} Many researchers have explored how to build runtime systems that support flexible pipeline parallelism. One early attempt was torchGPipe \cite{kim2020torchGPipe}, which implemented GPipe-style pipeline parallelism based on PyTorch's eager execution framework. DeepSpeed \cite{10.1145/3394486.3406703} and Megatron-LM \cite{megatron} also support pipeline parallelism in their training engines using a 1F1B style. Fairscale \cite{FairScale2021} provides experimental support for pipeline parallelism using PyTorch RPC \cite{paszke2019pytorch} with RemoteModule and message passing. Sagemaker \cite{karakus2021amazon} offers a flexible parallel programming interface that allows for partitioning of arbitrary models and pipeline parallelism with minimal code changes. In Hanayo, we have designed and implemented a runtime system that is decoupled from the pipeline parallel scheduling algorithms. This allows the runtime system to support any of the current pipeline parallelism scheduling algorithms and provides an interface for users to customize the scheduling algorithm.

\textbf{Techniques for Large Model Training.} The field of large model training can be categorized into two main areas: 1) \textit{Hybrid Parallelism:} Megatron-LM \cite{megatron} combines tensor parallelism and pipeline parallelism for large model training, utilizing tensor parallelism within nodes and pipeline parallelism between nodes. ColossalAI \cite{bian2021colossal} proposed other tensor parallel methods, such as 2D parallelism \cite{xu2021efficient} and sequence parallelism \cite{li2021sequence}, which can be combined with pipeline parallelism. 2) \textit{Memory Saving Techniques:} To reduce memory consumption during training, researchers have developed techniques such as activation checkpointing \cite{chen2016training, kirisame2020dynamic}, mix precision training \cite{micikevicius2017mixed}, and the ZeRO optimizer \cite{rajbhandari2020zero} proposed by DeepSpeed \cite{10.1145/3394486.3406703}. ZeRO and other works also support tensor offload \cite{ren2021zero, rajbhandari2021zero, patrick2023fang}, which allows for the use of CPU memory or even NVMe storage. These techniques are independent of pipeline parallelism and can be combined to improve large model training.

\section{Conclusion}

Hanayo introduces a novel pipeline parallelism scheduling approach that decouples the relationship between the number of stages and devices, leading to higher throughput, as well as an efficient framework that enables communication and computation overlap for any pipeline scheme. Scaling and adaptivity experiments demonstrate Hanayo's capability to handle various models, diverse computational environments, and different scenarios, such as large-scale training and fine-tuning. We believe that our proposed method will benefit both academia and industry through its efficiency and high usability.

\begin{acks}

\textbf{LIU} designed and wrote the system and the experiments, wrote the method and experiment-related parts in the paper, and participated in improving the method. 
\textbf{CHENG} designed and proposed key algorithms and structures, participated in paper writing, the improvement of codes implementation, and the experiments. 
\textbf{ZHOU} participated in the improvement of code implementation, experimental design and implementation, theoretical formula derivation, and article writing. 
\textbf{YOU} supervised this work and gave important insights.

This work used the Lonestar6 Cluster from TEXAS ADVANCED COMPUTING CENTER(TACC) and the cloud service of Tencent. We would like to thank them for their outstanding computing resource and professional service. ChatGPT was utilized to polish some of the texts in this paper. Yang You's research group is being sponsored by NUS startup grant (Presidential Young Professorship), Singapore MOE Tier-1 grant, ByteDance grant, ARCTIC grant, SMI grant and Alibaba grant.

\end{acks}


\bibliographystyle{ACM-Reference-Format}
\bibliography{sc23.bib}




\end{document}